\documentclass{article}

\usepackage{graphicx,amsmath,amssymb}

\newcommand{\be}[1]{\begin{equation}\label{#1}}
\newcommand{\ee}{\end{equation}}

\begin{document}
\begin{center}{\Large A queueing theory description of fat-tailed price
   returns in imperfect financial markets} \end{center}

\begin{center}

\underline{HARBIR LAMBA}\footnote{Department of Mathematical Sciences,
George Mason University, MS 3F2, 4400 University Drive,
Fairfax, VA 22030 USA}\\

\end{center}

\begin{abstract}

In a financial market, for agents with long
investment horizons or at times of severe market stress, it is often
changes in the asset price that act 
as the trigger for transactions or shifts in investment position. 
This suggests the use of price thresholds to simulate agent
behavior over much longer timescales than are currently used in models
of order-books. 

We show that many phenomena, routinely ignored in
efficient market theory, can be systematically introduced into an
otherwise efficient market, resulting in models that robustly
replicate the most important stylized facts.

We then demonstrate a close link between such threshold models and queueing
theory, with large price changes
corresponding to the busy periods of a single-server queue. The distribution
of the busy periods is known to have excess
kurtosis and non-exponential decay under various assumptions on the 
queue parameters. Such an approach may
prove useful in the development of mathematical models for rapid
deleveraging and panics in financial markets, and the 
stress-testing of financial institutions.\\
\mbox{}

\noindent PACS numbers 89.65.Gh 89.75.Da

\end{abstract}

\section{\label{intro}Introduction}

Economists and physicists have uncovered
seemingly universal statistical properties of real markets, many of
which deviate from those predicted by assumptions of market efficiency.
These are  often
referred to as the `stylized facts' \cite{c01,ms00} and there now exist
many different heterogeneous agent models (HAMs) that can replicate the most
important ones: the lack of correlations in the price-returns at all but
the shortest timescales; apparent power-law decays for the frequency of 
large magnitude
price changes ; and volatility clustering. 
It is not our intention to review the vast HAM literature here (a valuable
overview can be found in \cite{szsl07}) but many of the models
suffer from one or more of the following (related) problems. 

Firstly,
they tend to be constructed without due regard to the actual
process by which agents arrive at their chosen course of
action. This makes it difficult to  
argue why one model should be preferred over another which 
 in turn
makes it harder to convince orthodox economists to take any of 
them seriously. 

Secondly, at the level of individuals, 
many of the 
recent findings of behavioral economics
\cite{kt74,ds93,kr95,bt03} are ignored. Similarly, larger-scale market
structures and institutions may have rational-but-complex
motivations and perverse incentives that are also overlooked.

A third common problem is that agents are treated as
Markovian in the sense that their recent past does not influence their
future behavior. It occurs in those models that, for example,
probabilistically switch agents 
between investment positions or trading strategies\cite{am07,alw08}.

Finally, many models are sensitive to the size of the
system and when the number of agents
$M \rightarrow \infty$ some of the stylized facts, such as excess kurtosis,
 can even disappear
altogether.
 A frequent cause of this modelling issue is related to
the Markovian modelling of the agents mentioned above --- the Central
Limit Theorem and The Law of Large Numbers remove any endogenous 
fluctuations in the continuum limit.

Previous work \cite{cgls05,cgl06,cgls07,ls08} has shown that the use
of 
price thresholds to trigger
agent activity bypasses these problems while allowing for the modelling of
multiple real-world phenomena in a plausible and consistent
manner. Furthermore, an efficient market (where the price follows a
geometric Brownian Motion) exists as a special case within this framework
which makes possible a systematic study of the ways in which irrational
behavior and other `imperfections' may perturb such hypothetical solutions.
 
The two main contributions  of this paper are, firstly, to extend and
better justify the moving 
threshold models first introduced in \cite{ls08} and, secondly, to
show that queueing theory 
\cite{bps04} may provide both insights and analytical tools for
studying the  
fat-tailed price returns generated by such threshold models. 
This is because the largest price changes are caused by cascades of
buying or selling and their distribution can be reinterpreted
as the distribution of the 
{\em busy-period} of a single-server queue --- the length of
time for which a queue exists after it has begun. There are various established
results concerning the excess kurtosis and non-exponential decay of
this random variable under very general
assumptions on the arrival rate, departure rate and service time of
customers. While none of these extant results from queueing theory apply
precisely to the more 
complicated situation in market models, the correspondence is close enough to
suggest a common underlying mathematical explanation for the presence of
power-laws and fat-tails in both types of system.

The paper is organized as follows. Section~\ref{sec2} provides a
better motivation for  the moving 
threshold models first introduced in \cite{ls08}. In particular, a
separation of timescales argument is used to justify the main 
modeling assumptions. It is then shown how the rules governing the
threshold  dynamics  can mimic
many phenomena that are neglected in efficient/rational market models.  
It is especially interesting to introduce those very simple rules and
behaviors that induce 
coupling  between agents' trading strategies into an otherwise
efficient market. Power-law fat-tailed price returns and volatility clustering
consistent with the stylized facts are
robustly generated.

The class of models in Section~\ref{sec2} assumes a separation of timescales and
operates over long time periods. However, the fast cascade processes
responsible for the largest price changes can also be
viewed as a stand-alone model for very rapid price-deleveraging or
market panics,
say. Thus, in  Section~\ref{sec3} we consider such cascades and, after
a brief overview of queueing
theory, demonstrate the very close connection between the two. 
Standard queueing theory results concerning
the distribution of the busy-period of single-server 
queues then suggest a 
novel, but incomplete,
explanation of the fat-tailed nature of price returns in financial
markets.

\section{\label{sec2} A Threshold Model Over Long Timescales}

\subsection{A continuous-time model}

 We consider a market with asset or index price $p(t)$ at (continuous)
time $t$ and introduce a separation of timescales. 
Information arrives continuously and results in instantaneous 
price
changes that are implemented by `fast' agents who are primarily motivated
by such new information rather than price (these agents will not be
simulated directly in the time-discretized version since they operate over very 
small timescales). 
There are also, however, $M$ 
 `slow' agents, who are
motivated primarily by price changes, and act over much longer
timescales (typically weeks or months). 
Each of these $M$ agents can be either own
 (the state $+1$) or not own 
(the state $-1$) the asset at any given moment. 

\begin{figure}[ht]
\vspace{0.2in}
\centering
\includegraphics[scale=0.5]{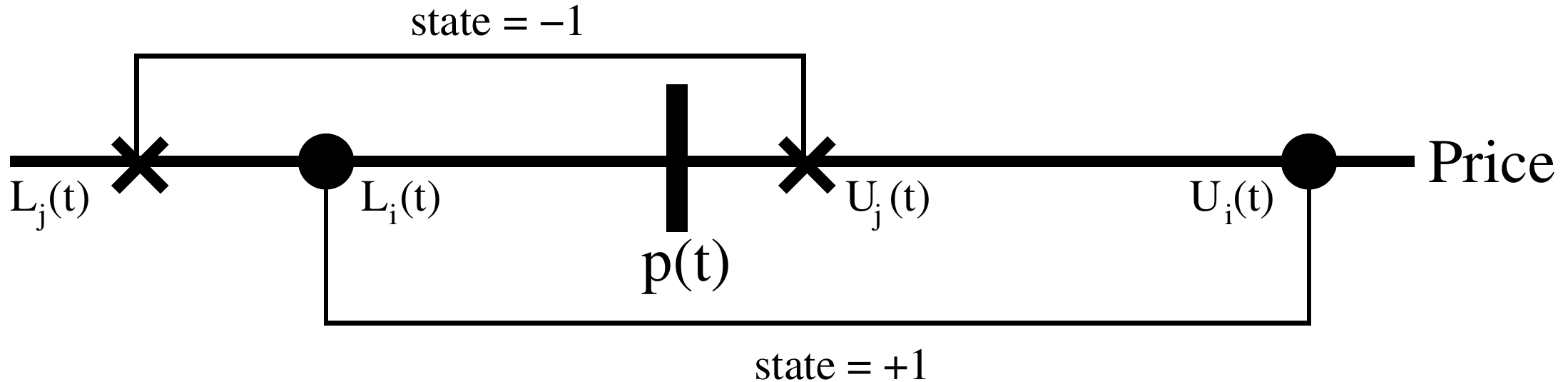}
  \caption{A representation of the model showing two agents in
    opposite states. Agent $i$ is in the $+1$ state and is represented by the
  (interval between) the two circles, and  agent $j$ is in the $-1$
  state and is represented by the two crosses.}
  \protect{\label{fig1}}
\end{figure}

At time $t$ the $i^{\rm th}$ slow agent is represented by its state, $\pm
1$, and an interval $I_i(t) = [L_i(t), U_i(t)]$ where $L_i(t) \leq p(t)
\leq U_i(t)$ (see Figure~\ref{fig1}). Whenever $p(t)$ crosses either
endpoint of this
interval, agent $i$ is deemed to be no longer comfortable with her
current investment position, switches states,  and the interval
$I_i(t)$ is updated so that $p(t)$ is again an interior point. Also,
the action of switching causes a small jump in the price caused by the
change in buy/sell pressure. Thus
at time $t$ the system is represented by the price $p(t)$, the states
of the agents, and $M$ closed intervals each of which includes the
value $p(t)$.     

We shall refer to the endpoints of the intervals as {\em
  thresholds} and the behavior of the system will be defined by
rules governing both the dynamics of $p(t)$ and the thresholds.
Before describing the kinds of rules that can be incorporated into
such a framework a few general remarks are in order.

The above model bears a resemblance to order-book
models \cite{s08} that attempt to describe how trades are cleared. In such
models, the price typically moves along the positive real line and
potential buyers and sellers are matched when the price is acceptable
 to both parties. However, the
differences are more profound than the similarities. Order-book
models are concerned with very short timescales and {\em how} trades
are executed. Here we are concerned with much longer timescales and {\em
  why} trades are placed --- it is implicitly assumed that 
the market is liquid (the fast sellers provide the necessary pool of
liquidity) and all trades
can be executed without explicitly matching
buyers and sellers.

\subsection{A discrete-time version}

 We  require  time-discretized models that are 
suitable for computer simulations.
First a timestep $h$ is chosen, typically less than one trading day.
The state of the $ i^{th} $ slow agent over the $ n^{th} $ time
interval is represented by
  $s_i(n) = \pm 1 $.
The price of the asset at the start of the
$n^{\rm th}$ time interval is $p(n)$ and for simplicity the system is
 drift-free so that $p(n)$ actually corresponds to the price relative to
the risk-free interest rate plus equity-risk premium or
 the expected rate of return. We do  not assume that
 agents are of uniform size (in terms of their trading positions) and
 thus weight the $i^{\rm th}$ agent by its size $w_i$ and define $W =
 \sum_{i=1}^M w_i$.
A key variable is the {\em sentiment} defined as 
the weighted average of the states of all of the $M$ investors
\begin{equation} \sigma (n) = \frac{1}{W}  \sum_{i=1}^M
  w_i s_i(n).\label{sigma} \end{equation} 
and $\Delta \sigma (n) = \sigma (n) - \sigma(n-1)$.

The price $p(n)$ evolves according to the rule 
  \begin{equation}
p(n+1)=p(n) \exp\left(\left(\sqrt{h}\eta(n)\right)
f(\bullet ) -h/2 + \kappa \Delta\sigma(n)\right)\label{price1}
\end{equation}
where $\kappa > 0$ and $\eta(n) \sim {\cal N}(0,1)$ represents the exogenous
information stream. 
Note that when $\kappa = 0$ and $f\equiv 1$ the price follows a
geometric Brownian motion (the term $-h/2$ is the drift correction
required by It\^{o} calculus). 
The function $f(\bullet )$ allows the effect of new
information on the marketplace to vary 
and is discussed
further below but note that if $f \equiv 1$ then the pricing formula
\eqref{price1} is simply a geometric Brownian motion modified by a
linear supply/demand correction due to the sentiment of the slow
agents.

The rules governing the dynamics of the thresholds are implemented as follows.
The thresholds for each agent change (usually slowly) between switchings
and correspond to that agent's
evolving  {\em strategy}. 
If, at the end of a timestep, an agent's interval is crossed by the
price, then that agent switches and the corresponding change in
$\sigma$ will feed into the price at the next iteration. This choice
of synchronous updating is made for two reasons --- it is 
computationally convenient but  also reflects the fact that even
after a slow trader decides to react, there is likely to be some
 delay
in effecting the trade (unlike the fast traders).
In the simulations that follow, $h$ is chosen to correspond to 1/10 of a
day and then the daily price returns are computed).

As general as the above framework is, it
can be made more realistic by allowing agents to own differing
amounts of the asset, and perhaps even shorting. This can be achieved
by assigning an (evolving)  weight value to each threshold (upper and lower,
separately) such that when that threshold is breached the agent buys
or sells so as to own that amount of stock. However useful this may
prove to be in the future, it will not be considered here. 
 
It is important to note that the model has two fundamental, and
essentially separable, components --- one  
governs the motion of the price $p(n)$, given by
\eqref{price1}, and the other governs the motion of the
thresholds. Or,
equivalently, one set of rules describes the fast agents and the other 
describes the slow ones.
We now discuss  which phenomena can be incorporated 
into such a model. 

Firstly, in the pricing formula \eqref{price1} the law of supply and
demand is reproduced for $\kappa >  0$ --- increasing/decreasing $\sigma$ causes
the price to rise/fall. Its magnitude (relative to $f$) determines the
extent to which price changes are determined by changes in sentiment
versus the arrival of new information. If the function $f \equiv 1$
then the fast traders are accurately and instantly converting information
into price changes.   
 We posit, however, that this is not necessarily the case. For example
during times of extreme sentiment $f(\bullet )$ may be greater, perhaps due to a
surplus of speculators
\cite{b99} or excessive attention being paid to information
in an environment where market conditions are perceived to be due for
a correction of some kind.
This mechanism is undoubtedly too
simplistic but nevertheless tying volatility to sentiment in this way results in
realistic volatility clustering \cite{ls08}. Or one can instead
introduce an  explicit dependence of $f$ upon $n$ to create changes
in market 
conditions with time or to use a stochastic volatility model.  
And of course it is still being assumed
that the information stream is Gaussian and uncorrelated with itself 
--- weakening these highly unrealistic assumptions
provides additional mechanisms for volatility clustering. In
actual  markets,
all of these mechanisms are probably present to some degree or
another, and will be considered in more detail elsewhere.

\subsection{Introducing market `defects' using  thresholds}

The focus of this paper 
is on the fat-tailed price returns, and previous results
\cite{cgls05,cgls07,ls08,ls06} suggest that these are caused by
the thresholds and their  dynamics rather than the pricing mechanism.

Let us start by assuming that $f\equiv 1$; that the threshold
distributions of the agents are perfectly mixed along the priceline 
at $n=0$; and
that the threshold rules for agents in either state are identical and
simply geometric Brownian motion. Then $\sigma $ will not move
away from zero since equal numbers of slow agents are switching
between the two states and so $p(n)$ reduces to a geometric
Brownian process.  Thus this special case corresponds
to a weakly-efficient market \cite{f70} 
with no possibility of predicting future prices
from prior ones. 

This base model is also very closely related to the
concept of {\em rational expectations} (in the sense of \cite{m61}) 
 which relies upon
the assumption that the predictions and expectations of all the agents
are, on average, correct.  In other words, there are no systemic
biases or dependencies 
between agents' strategies which are treated as Gaussian
fluctuations around the `correct' one\footnote{This 
  assumption is used in macroeconomics to justify the use of a single
  `representative agent' to model an entire economy.}. 
Thus, interpreting the random
motions of the slow agents' thresholds  in the above model as, in
fact, being the 
result of arbitrarily complex computations to maximize 
individual utility functions gives a model that is, practically
and philosophically, very close to the neoclassical paradigm.

But, by allowing for a wider range of threshold dynamics, we can replicate
many different effects observed in real agents and real markets ---
and in doing so violate
the above assumptions of independence and lack of systemic bias in
agents' strategies. 

Suppose first that agent $i$ 
is perfectly rational then the values $L_i(n)$ and $U_i(n)$ represent
the cumulative effect of rigorous market
analysis and future expectations. Thus the price interval
$[L_i(n),U_i(n)]$ defines that agent's (conscious, or algorithmic in
the case of a trading program) 
strategy.
Or in the case of a less-than-perfectly-rational individual the price interval
still represents a de facto strategy, but one that the agent herself
may not be fully aware of. The rules governing the dynamics of the
threshold values may be as complicated as desired, simultaneously incorporating
amongst many other things: rational strategies based upon optimization
of utility functions or econometrics or technical analysis;
past performance; adaptive heuristics; inductive learning; new
information; tax issues; price data from alternative investment options;
margin calls; perverse incentives; recent market volatility (perhaps
then influencing future volatility); 
psychological effects; the weather; herding; imitation within a subset of
closely-networked agents;  
and all the key findings of behavioral economics. 

Thus another working assumption 
is that all such effects are cumulative and
can be applied to a single pair of thresholds --- some effects will move
a particular threshold inwards towards the current price (making it more likely
that that threshold will be breached and the agent will switch) 
while others will
move it outwards. In other words we are hypothesizing that a slow
agent's 
past experiences, present state and future expectations can be
condensed into two price
values, one on either side of the current price, together with the
current state of the agent. Information about the
agent's state, threshold values and threshold dynamics are all carried
over from one timestep to the next.

Thresholds are especially well-suited for incorporating 
some important aspects of agent behaviors and motivations in
a very natural way.
 By ensuring that they are reset away from the
current price after a switching, agents cannot  switch
arbitrarily often 
(which will be the case in the presence of non-negligible transaction
costs). Similarly included is  the `anchoring' phenomenon 
whereby the price 
at which the agent last traded influences the value they place upon
the asset and thus the price at
which they next trade. In a similar vein, chartists and technical
analysts have developed many observational rules to help them predict
future market performance. A particularly simple kind concerns the
existence of `resistance levels' which, if breached, indicate a
further price move in the same direction. If such an effect does
indeed exist then it can be applied to the threshold dynamics of a
subset of the agents (or it may even arise as a natural
consequence of other rules).  Taking this one step further, a subset
of agents may be influenced by important-sounding numbers, 
such as
10000 on the DJIA, and consciously-or-not have a threshold at around that
value\footnote{Or assume that other, less rational, agents will be 
affected by it (as in Keynes' beauty contest) 
and so assign a significance to it
  themselves. In either case, the effect is the same!}. 
As another (extreme but not uncommon) example consider an individual who bought
a dot-com stock at the top of the tech-bubble and then held it all
the way down to 
zero. In such a case, that agent's lower threshold decreased (due to 
loss aversion) even faster than the price while the upper threshold 
didn't
decrease fast enough and so the stock was never sold. At the more
rational
end of the spectrum,
a rational investor or computerized trading program (fundamentalist trader)  
that believes the current
price to be too high, with no other complicating factors, 
will enter state $-1$ (if necessary) and then
can be mimicked by setting
the
lower threshold at some point below the appraised price level and the
upper level arbitrarily high.

However the phenomenon that appears to be most important in the
generation of fat-tails within threshold models is {\em herding}, 
whereby there is a tendency for (rational or otherwise) agents in 
the minority position to switch and join the majority. 
 This can be incorporated into threshold dynamics in a
very simple manner --- agents in the minority position have their
thresholds move inwards towards the current price thus making them
more likely to switch and join the majority (unless the majority
state changes first). Different agents have differing herding 
propensities that are
reflected in the rate at which their thresholds move together.

Herding may be initiated, for example, by a widespread misconception about   
the present or the future (such as house prices never going down) or some
systemic asymmetry\footnote{For example the perverse incentives induced by
inappropriate fee structures, or the moral hazard that occurs when the
risk of a certain investment position is reduced, removed or
transferred.}  that causes agents to prefer one of 
the states over the other. Herding then provides a plausible positive-feedback
mechanism.
However, it must be emphasized that for many agents herding is a rational, not
irrational,  phenomenon.  
Indeed herding in the natural world is an effective survival strategy
and the same is 
true in financial markets (as well as being a trading strategy in
itself, often referred to as momentum trading). Professional investors risk
losing their jobs, bonuses and/or investment capital if they deviate too far
from the mean in what turns out to be the wrong direction. Thus there
is a strong motivation
to play safe and  `chase the average'. This naturally (and
rationally from the 
point of view of the agents themselves) also leads to herding.

\subsection{ Numerical Simulations}  \label{num}

It is not the purpose of this paper to provide a detailed numerical
investigation since results from a similar model (using two pairs of
static thresholds for each agent) are compared against the stylized
facts in \cite{ls06} and further numerical results for a  moving 
threshold  model can be found in
\cite{ls08}. However, for completeness,
we provide enough details for
reproduction of the numerical results and highlight the main findings of 
previous studies.

The model is kept as simple as possible and so
apart from including herding and a volatility function $f(\sigma)= 1 +
2 |\sigma| $ 
we assume that agents thresholds are reset after a switching from a 
specified random distribution.

The timestep $h$ is defined in terms of the variance of the
external information stream. A daily variance in price returns of 0.6--0.7\%
implies that $h$ of 0.000004 should correspond to approximately 1/10 of
a trading day. The results of 10 consecutive timesteps are then combined to give
the daily price return. 

The parameter $\kappa = 0.1$ and all agents have equal weight.
The reset thresholds after a switching at price $P$ are $[L_i,U_i] = [P/(1+Z_L),
P(1+Z_U)]$ where $Z_L,Z_U$ are each chosen from the uniform
distribution on the interval $ [0.05,0.25]$, corresponding to price moves
in the range 5--25\%. The model is very robust to changes in these 
parameters and in the absence of further information they are chosen to
be as simple as possible.

The thresholds of all agents are subject to a random, driftless, 
component governed by a quantity $\delta = 10^{-8}$.
Herding is introduced by supposing that for agents in the
minority position only
$$ L_i(n+1) = L_i(n) + C_i h |\sigma(n)| + {\cal N}(0,\delta), \quad 
 U_i(n+1) = U_i(n) - C_i h |\sigma(n)| + {\cal N}(0,\delta)$$
while for those in the majority
$$ L_i(n+1) = L_i(n)  + {\cal N}(0,\delta), \quad 
 U_i(n+1) = U_i(n) + {\cal N}(0,\delta).$$
 Note
that the drift in the position of the thresholds is proportional to
the length of the timestep and the magnitude of the sentiment. The
constant of proportionality $C_i$ is different, but fixed, 
for each agent and chosen from the uniform distribution on
$[20,100]$. This range of parameters corresponds to a herding tendency
that operates over a timescale of a few months or longer. 

\begin{figure}[ht]
\centering
\includegraphics[scale=0.3]{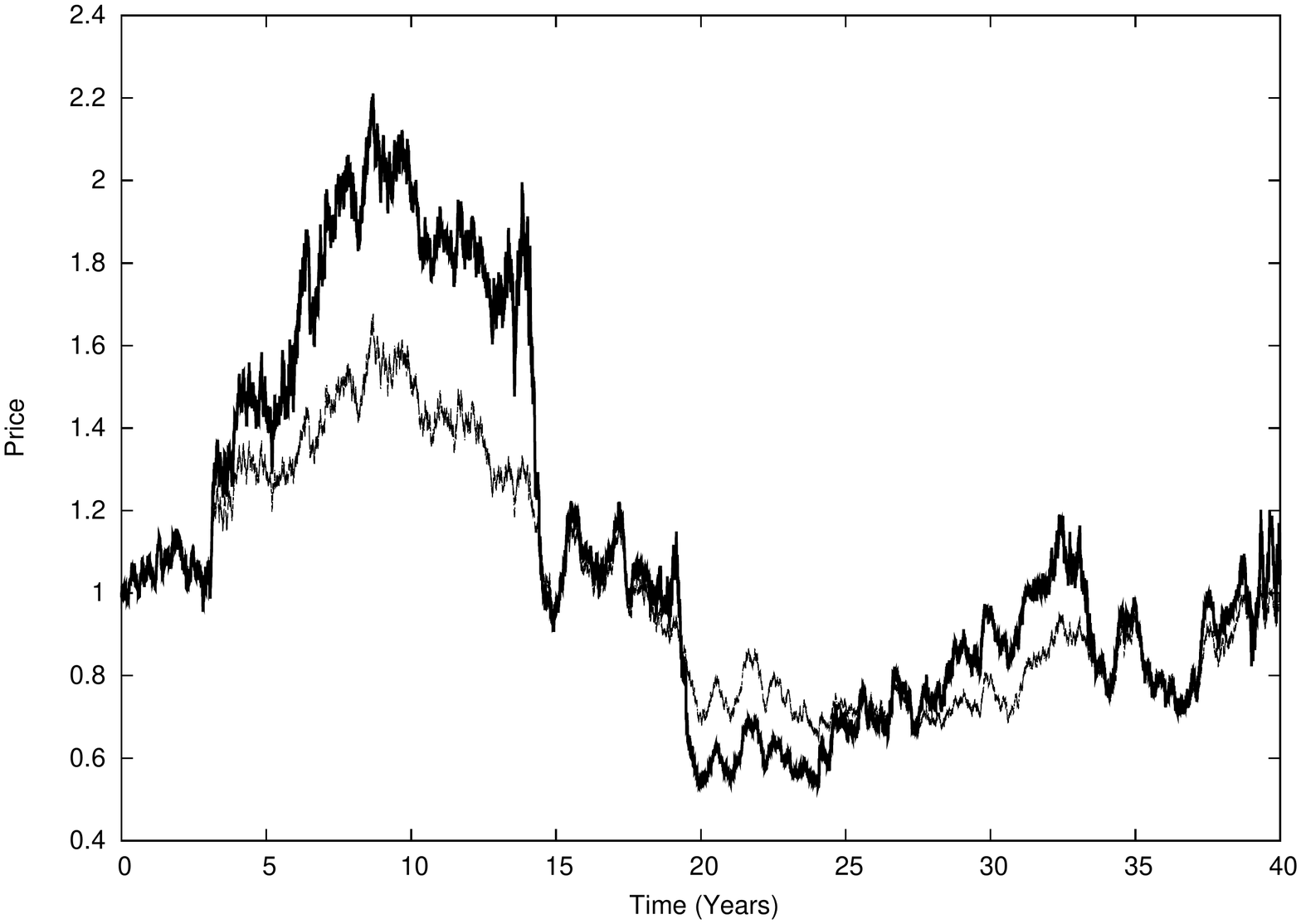}
  \caption{ Asset price of a simulation over 40 years with 100000 agents.}
  \protect{\label{fig2}}
\end{figure}

Figure~\ref{fig2}
shows the price output of the model (the more volatile curve) against 
 the
`efficient' pricing (less volatile)
obtained from (\ref{price1}) by setting $\kappa = 0$ and $f(\sigma)
\equiv 1$ with $M=100000$. 

\begin{figure}
\centering
\includegraphics[scale=0.3]{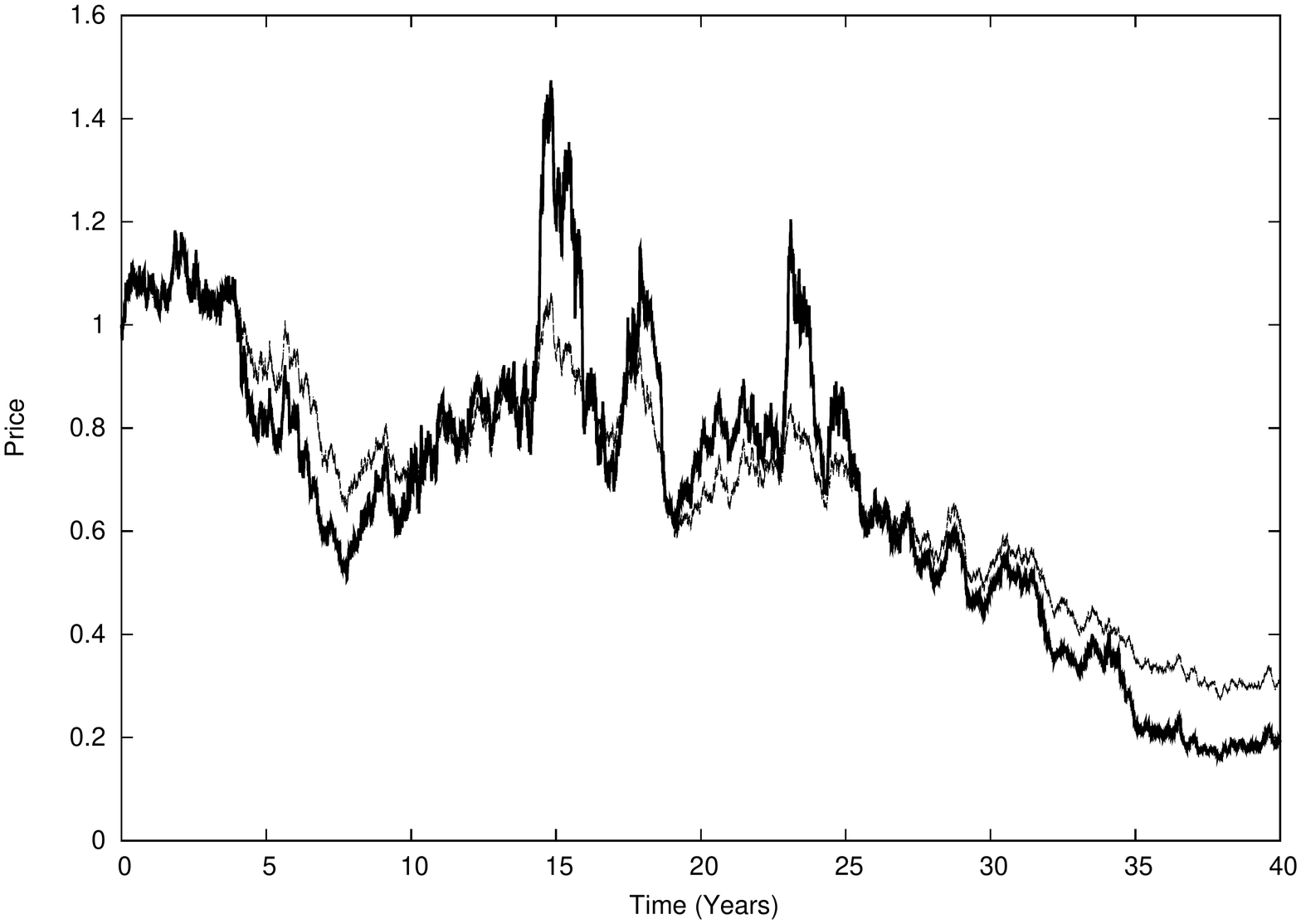}
  \caption{ Asset price of a simulation over 40 years with 1000 agents.}
  \protect{\label{fig3}}
\end{figure}

\begin{figure}
\centering
\includegraphics[scale=0.3]{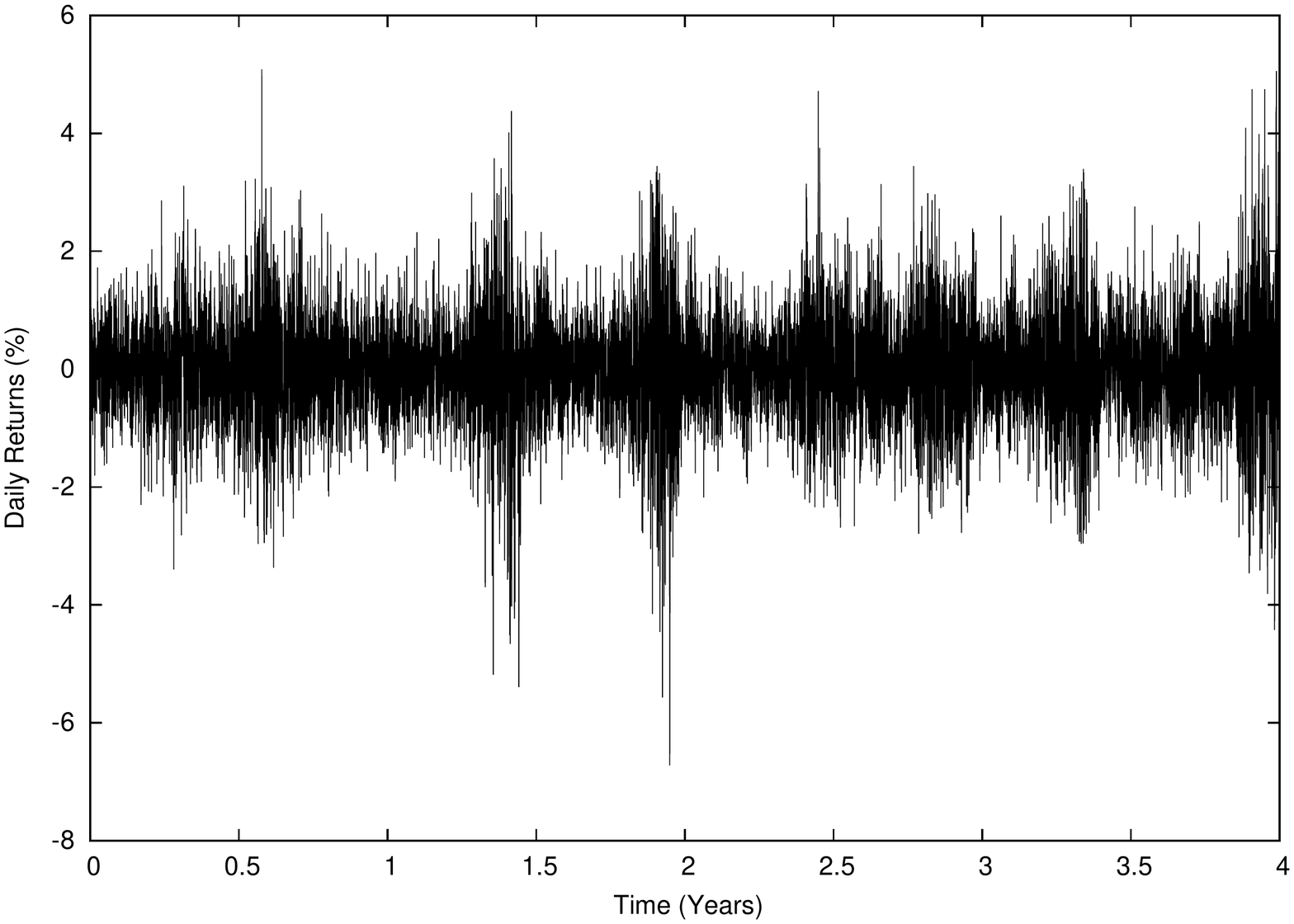}
  \caption{ Daily price returns for Figure 2.}
  \protect{\label{fig4}}
\end{figure}

 Figure~\ref{fig3} shows a simulation with $M=1000$ but all other 
parameters unchanged, emphasizing that the 
results do not depend critically upon the system size while Figure~\ref{fig4} shows the daily percentage returns for the simulation in 
Figure~\ref{fig2}. 

\begin{figure}
\centering
\includegraphics[scale=0.3]{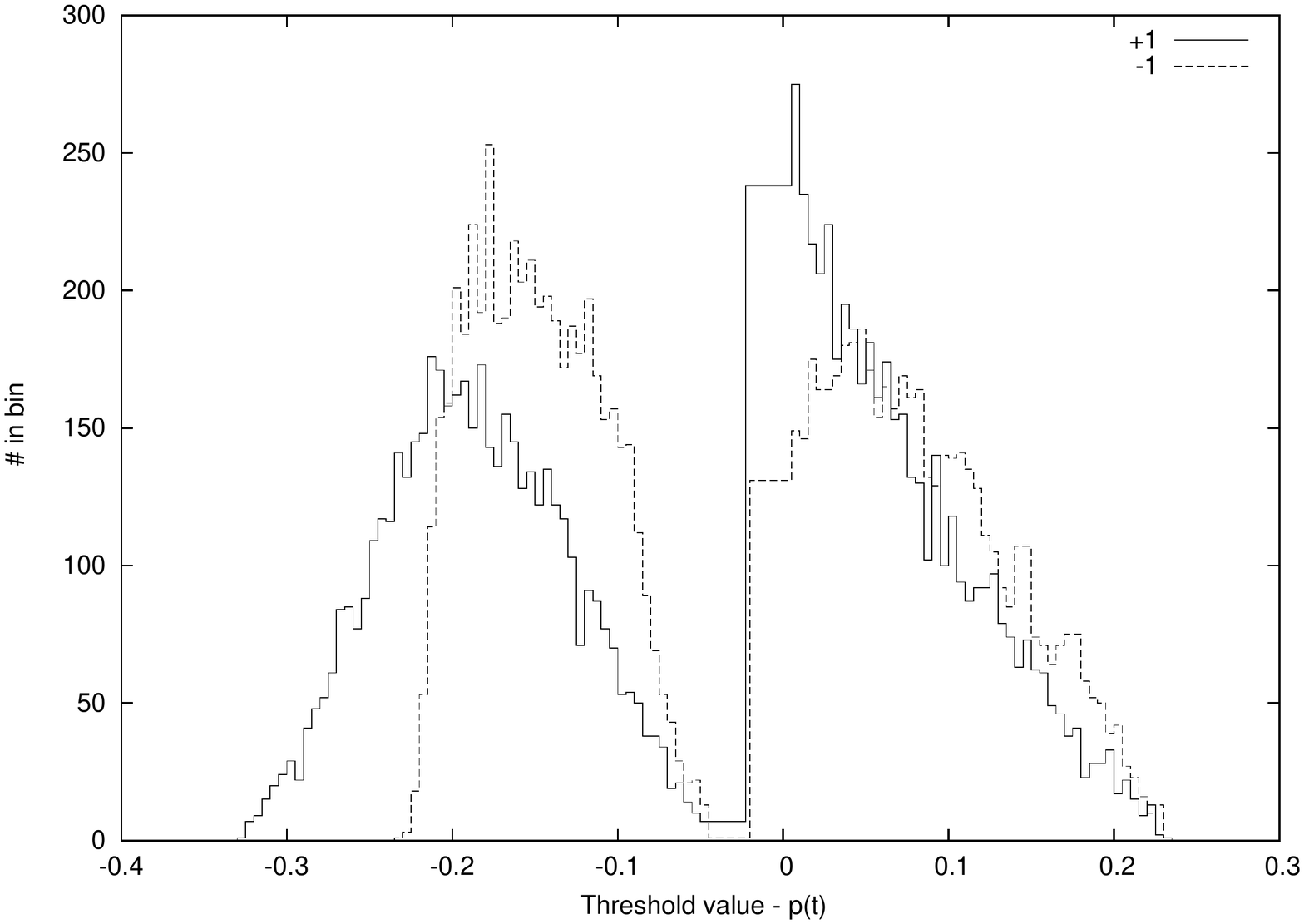}
  \caption{ The density of the thresholds along the price axis
 for agents in each of the
    two states. The
    difference between 
    the two distributions is a `memory effect' of the prior behavior
    of the system and affects the future evolution.}
  \protect{\label{fig5}}
\end{figure}

Measurements of power-law exponents, similar to those carried
out in \cite{ls06}, provided estimates close to those observed in
analyses of price data from real markets for the tail of the price
returns and the decay of the volatility autocorrelation function,
typically in 
the range $[2.8,3.2]$.

In
the absence of herding, i.e. when $C_i = 0 \; \forall i$, then,
provided that the initial states $s_i(0)$ of the agents are
sufficiently mixed, $\sigma(n) \approx 0 \; \forall n$  and $p(t)$
always remains close to the efficient market price. Such a model is 
both practically and philosophically recognizable as the neoclassical
notion of an efficient market --- agents trade due to differing
expectations but the averaging procedure inherent in the rational
expectations assumption is valid and no mispricing occurs.
However, 
Figure~\ref{fig5} is a snapshot of the system for the simulation of
Figure~\ref{fig2} (herding included) at a point in time when $\sigma\approx 0$. The two displayed
histograms are the density of thresholds for agents in the two states $+1$ 
and $-1$ plotted separately. The main point to note is the discrepancy between 
the two density functions which indicates that as the system evolves and 
the price fluctuates, $\sigma$ will move away from 0 because
  the agent thresholds
are not perfectly mixed.

Finally we note that in the above simulation there are different
causal mechanisms for the volatility
clustering and fat-tails. If $f\equiv 1$ (i.e. the fast agents are 
correctly interpreting new Gaussian information) then herding induces 
fat-tails but there is no volatility clustering. Thus the systemic
bias of the slow agents cause fat-tails while 
the imperfect interpretation of information by the fast agents results 
in volatility clustering\footnote{However, more sophisticated
  threshold dynamics for the slow agents that are also functions of
  recent price volatility may provide a further mechnism for
  volatility clustering}.
 
\section{\label{sec3} A Queueing Theory Description of Price Cascades over Short Timescales}    

\begin{figure}
\centering
\includegraphics[scale=0.3]{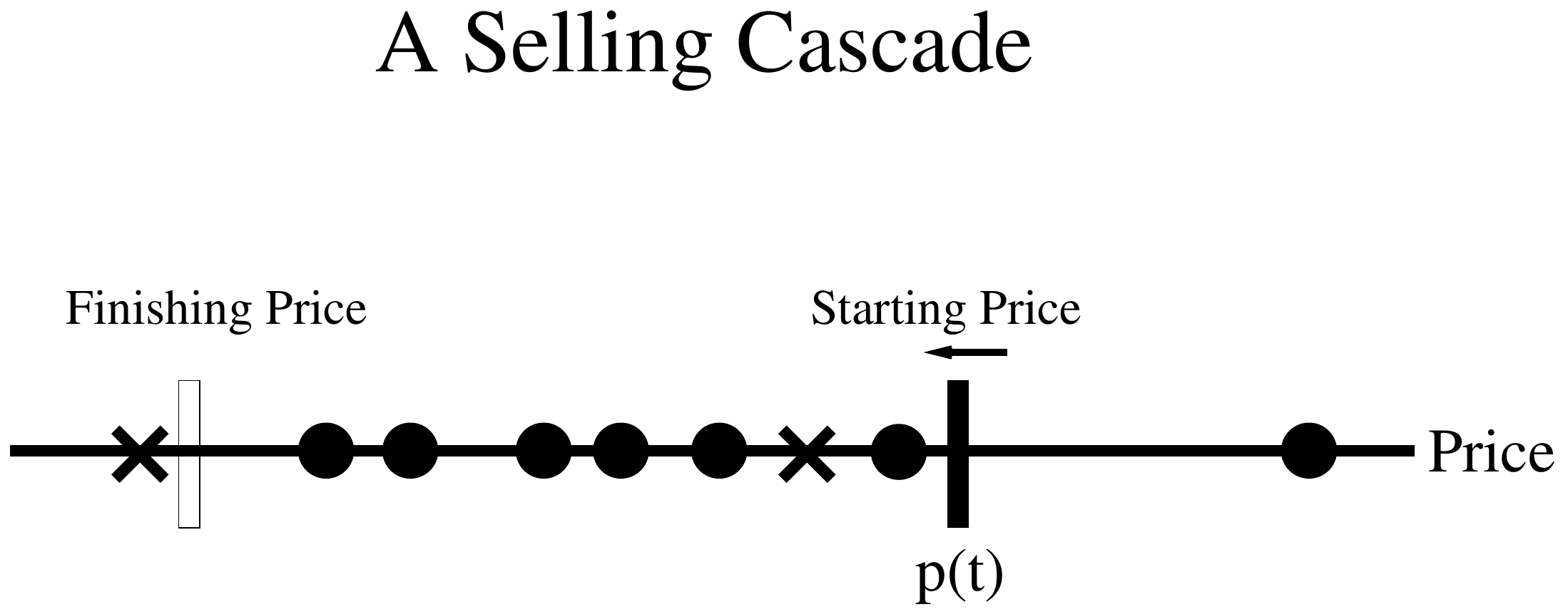}
  \caption{The start of a selling cascade. As in Figure~1, crosses and
    circles 
    represent thresholds of agents in the $-1$ and $+1$ states respectively.}
  \protect{\label{fig6}}
\end{figure}

We are interested primarily 
in the tail-distribution of price returns. Since
the information stream is modelled as Gaussian, extreme price moves are
due to cascades of buying or selling affecting \eqref{price1} with $\kappa > 0$. Without loss of generality we shall
consider a selling cascade. Figure~\ref{fig6} shows the start of
such  a possible selling 
 cascade within the threshold model of Section~\ref{sec2}. The circles
and crosses to the left of the current price indicate the positions of the
lower thresholds of the slow agents.

There are four points to consider before proceeding. Firstly, Figure~\ref{fig6}
can act as a stand-alone model of a rapid deleveraging process. As the price
falls and passes the lower thresholds of agents who are in the $+1$
state,  
they switch to the $-1$ state 
pushing the price down further (if $\kappa >0$) 
and triggering other agents to sell and so on (note that these lower 
thresholds may in reality be the pricing points at which margin calls are
activated).  
Secondly, under the most
extreme market conditions, the distinction between fast and slow agents made in
Section~\ref{sec2} may be invalid as the amount of new information entering the
system is negligible and all the agents are motivated by price
changes. In this case
$M$ will be equal to the total number of agents, not just the number of slow 
ones, and the model more closely resembles an order book. Thus the
link to queueing theory outlined below may also be useful in the context of
those models.

Thirdly, as shown in Figure~\ref{fig5}, the distribution of thresholds is a result of the evolution
of the system over a long period of time and will in general be highly non-uniform
over the positive real line. Finally, although we are supposing that
the cascade/relaxation process is instantaneous, in practice this is
certainly
not the case and agents' thresholds may move significantly between the
start
of the cascade and the end.  

We consider the
continuous-time version of the model outlined at the start of
Section~\ref{sec2} and it will also be convenient to introduce the
log-price $P(t) = \ln p(t)$. At the start of the cascade the price
 is fluctuating
due to the arrival of new information and the action of the fast
traders. Then at some value $P^*$ it equals one of the thresholds of agent
$i$ who is in the $+1$ state.  We now consider the cascade to be a relaxation
process that occurs instantaneously.
The log-price jumps downwards by an amount
$\kappa \Delta \sigma = 2 \kappa w_i/W$ as the agent switches and its
thresholds jump to a non-zero distance away. If there is no other
agent with its lower threshold in the interval $[P^*,P^*-2\kappa w_i/W]$ then
the cascade stops immediately. If there is such an agent, or agents,
then the cascade continues. During the relaxation process,  
there may be agents in the
opposite state caught in the cascade whose switching will act in the
opposite price direction and help to bring the cascade to an end. 
Once the cascade is over,
time restarts and $P(t)$ evolves under the action of the information
stream until another threshold is met.

Now consider the following scenario from queueing theory \cite{bps04}.
 A customer, named $i$, 
arrives at an empty single-server queue. If she is served before
another customer arrives then the {\em busy period} of the queue,
defined to be 
the length of time it is in existence, is simply the time taken
to serve her, call it $2\kappa w_i/W$. 
If on the other hand other customers arrive before she
is served then the queue continues until they are all served. However,
we must also allow for the possibility that some people in the queue
may decide to leave it before being served --- this is referred to in
the queueing theory literature as {\em reneging}. 

There exists an almost exact correspondence between these two situations. Price
in the market model corresponds to time in the queueing problem, the
size of the trade the agent wishes to make corresponds to the length
of time taken to serve that customer, and the overall price change
during the cascade maps to the busy period of the queue.  Agents in the
opposite state caught up in the cascade act as `anti-customers' whose
arrival in the queue causes them to cancel out with an agent, or agents, of the same total weight/service time.  This is of course 
equivalent to those agents deciding to renege and leave the queue.
A subtle, but negligible, difference is that in the market cascade an
agent $j$ in the opposite state getting caught up in the process
should correspond to a 
customer, or customers, of the same total `size' $w_j/W$ reneging from
the queue. 
But if $w_j$ is sufficiently large then there may not be enough total
weight in the queue for this to happen 
(i.e. 
 at the end of a selling cascade a small `bounce' may occur in the price but 
 customers who have already been served cannot leave a queue). 

Queueing theory has a standard notation for describing queues. A queue 
(without reneging) is
$X/Y/n$ where $X$ describes the distribution of arrivals, $Y$ the serving 
times and $n$ is the number of servers. Below we shall only consider 
cases where
$X,Y = M$ or $G$ and $n=1$. 
$M$ corresponds to exponentially-distributed arrival or service times 
 generated by a Poisson process while $G$ is a general distribution, 
usually under some assumption of finite moments. 
In the case of an $M/M/1$ 
queue the Poisson parameters for $X$ and $Y$ are $\lambda$ and $\mu$ 
respectively.

We now state some results from queueing theory concerning the
distribution of the busy period under progressively weaker assumptions.
While none of these results apply directly to price cascades under all
circumstances, they may be indicative of an underlying  explanation for
the apparent universality of sub-exponential decays in price returns 
\cite{ms00}.

First consider an $M/M/1$ queue without reneging. Using standard 
arguments \cite{bps04}, if  $\rho =  \lambda/ \mu < 1$ then 
 the busy period $\tau$ of the queue is 
finite with probability 1 and the entire distribution is given by
\begin{equation}{\rm Prob}(\tau \leq t) = \frac{\sqrt{\mu/\lambda}e^{-(\lambda+\mu)t} 
I_1 (2 \sqrt{\lambda\mu}t)}{t}\label{pd} \end{equation}
where $I_1(\cdot)$ is the first modified Bessel function of the 
first kind. Furthermore explicit formulae exist for the moments and 
\begin{equation}E[\tau^4] = \frac{E[Y^4]}{(1-\rho)^5}+ \frac{10\lambda
E[Y^2]E[Y^3]}{(1-\rho)^6}+  \frac{15\lambda^2
(E[Y^2])^3}{(1-\rho)^7}\label{m4}\end{equation} 
(generalizations for $M/M/1$ queues with various reneging assumptions
can be found in \cite{jd08}).
In fact \eqref{m4} holds for $M/G/1$ 
queues without reneging if $\rho = \lambda/E(Y) < 1$. The generalization
of \eqref{pd} to $M/G/1$ queues is provided by the 
{\em Takacs Equation}:\\
if $Y^*(s)$ is the Laplace-Stieltjes transform of the cumulative 
density function  of $Y(t)$  then
$\tau^*(s)$ satisfies
\begin{equation} 
\label{takacs} \tau^*(s) = Y^*\{s+\lambda -\lambda \tau^*(s)\}.\end{equation}

This functional equation establishes close connections 
between the tail of the service times and the tail 
of the busy period distribution \cite{mt80} (see also
\cite{z01,bdk04,jd08}). 
In particular, the tails of
the service time distribution (ie. the distribution of agent sizes)
and the tails of the busy period (price changes) obey the relationship
$$(1-\tau(t)) \sim (1-\rho)^{-\alpha -1}(1-Y(t)) \; {\rm as }\; t
\rightarrow \infty $$
where  $\alpha \geq 1$ and $ Y(t)$ is subexponential with $1-Y(t) =
\frac{L(t)}{t^\alpha} $ for some slowly-varying function $L$. 
Such results are of interest because of the observed power-law
distributions of sizes in many socio-economic and financial
systems. However it should be noted that threshold models can produce
approximate power laws even when the agents are all of uniform size
\cite{ls08} and so a general explanation of power-law price returns may
not require a power-law distribution for the agents' sizes.

The above results for $M/G/1$ queues without reneging 
may be directly applicable to smaller and intermediate-size
cascades where the presence of 
reneging (agents who are
switching in the opposing direction) can be approximated by
modifying the service time distribution and, in particular,
 decreasing $E(Y)$.
However, especially for larger cascades, the assumptions of constant  
arrival and  reneging rates are not appropriate. 
As an extreme case, consider a  selling cascade that 
initially has arrival rates
 corresponding to a supercritical process 
(and a queue that would become infinitely long with non-zero probability). 
Then the mechanism that ultimately stops the cascade is the re-arrival of
agents who have already been through the cascade once. 
For example, an agent
who was fortunate (or smart) enough to sell at the start of a selling 
cascade may eventually re-enter the process but making the opposite
trade, thus reaping an actual as opposed to hypothetical profit,
 and acting as a brake rather than an accelerator. Thus financial
cascades have a propensity to be self-limiting.  
Similar situations  within queueing 
theory arise by considering
the arrival and reneging rates to be  functions of time and/or queue
length \cite{sk82,ag63a,ag63b} but few general results exist
concerning the busy-period distribution. 

We end by commenting  that price-return tails 
 are an aggregate of many different cascades that almost 
certainly have differing queueing parameters. Thus a precise analysis
of the busy-period distributions may not be necessary to explain the
observed functional forms of the tails.

\section{\label{sec5}Conclusions}  

Even very simple threshold models appear to be capable of capturing
the most important statistical 
aspects of financial markets. They also better reflect
the incremental and history-dependent nature of decision-making
processes. Furthermore, since a (weakly) efficient market exists
as a special case, it is possible 
to systematically study the ways in which
removing various efficient market assumptions, as reflected in
the model, change the global behavior of the system. In this sense,
the framework presented in Section~\ref{sec2} can be considered as
a set of thought experiments that can be used to query the
assumptions behind much of modern finance and economics.

The numerics in Section~2.4 focussed upon herding as the
principle cause of deviation from an otherwise efficient market. This
is because many important real-world factors such as psychology,
bonus/commission criteria and other compensation practices, and moral
hazard all produce a similar effect; namely, that one side of a
trade or market position becomes more attractive than the other and
then herding provides a positive feedback mechanism.

It was then demonstrated that
queueing theory can potentially be applied to threshold models. 
It is unlikely that the cited
results, concerning the  
kurtosis and tail-distribution of the busy periods of
single-server queues, can be accurately applied to any real
market cascade. However, the analytical methods of
queueing theory may provide an explanation for the universality
seen in the tail statistics. Queueing theory may
also be useful for analyzing order-books and modelling market panics
and deleveraging cascades. 
 
Finally, in this paper the attention has been on the slow traders and
hence the threshold dynamics. However, such threshold models
can be fused, in a very intuitive way, with order-book models.
The resulting models would be computationally very intensive,
since they must simulate the
fast traders as well as the slow ones, but should be able to mimic many
different real-world effects across multiple timescales.

\bibliographystyle{plain}
\bibliography{cgl}

\end{document}